\documentclass[a4paper,11pt]{article}

\usepackage[utf8]{inputenc}
\usepackage[T1]{fontenc}
\usepackage{lmodern}
\usepackage{microtype}
\usepackage{amsmath,amssymb,amsfonts,bm,mathtools}
\usepackage{graphicx}
\usepackage{booktabs}
\usepackage{array}
\usepackage{multirow}
\usepackage{enumitem}
\usepackage{xcolor}
\usepackage{url}
\newcommand{\href}[2]{#2}
\usepackage[margin=1in]{geometry}
\usepackage{authblk}
\usepackage{titlesec}
\usepackage{fancyhdr}
\usepackage{caption}
\usepackage{tikz}
\usepackage{pgfplots}
\usetikzlibrary{arrows.meta,positioning,calc}
\pgfplotsset{compat=1.18}

\definecolor{MidnightBlue}{rgb}{0.02,0.12,0.38}
\definecolor{braneA}{RGB}{33,110,180}
\definecolor{braneB}{RGB}{200,70,60}
\definecolor{funnelgreen}{RGB}{55,145,95}
\definecolor{goldink}{RGB}{224,168,42}
\definecolor{purpleink}{RGB}{124,87,178}
\definecolor{softgray}{RGB}{244,244,248}

\titleformat{\section}{\Large\bfseries\color{MidnightBlue}}{\thesection}{0.6em}{}
\titleformat{\subsection}{\large\bfseries\color{MidnightBlue}}{\thesubsection}{0.6em}{}
\titlespacing*{\section}{0pt}{2.4ex plus .4ex}{1.2ex}
\titlespacing*{\subsection}{0pt}{2.0ex plus .3ex}{0.8ex}

\newcommand{\dd}{\mathrm{d}}
\newcommand{\Msun}{M_\odot}

\title{\bfseries Hidden-sector accretion and warped black-string seeds for high-redshift supermassive black holes}
\author[1]{Chunshan Lin}
\affil[1]{Faculty of Physics, Astronomy and Applied Computer Science, Jagiellonian University, 30-348 Krakow, Poland}
\date{}

\begin{document}
\maketitle

\begin{abstract}
The earliest massive black holes are often discussed in terms of heavy baryonic seeds, primordial black holes, or super-Eddington accretion.  We develop a different possibility: a compact object on a hidden donor brane forms a common five-dimensional horizon whose intersection with our brane is observed as a black-hole seed.  Donor-side matter accretes onto the common five-dimensional horizon, increases the  horizon radius and, consequently, also enlarges its intersection with our brane. As a result, the gravitational mass inferred by an observer living on our brane grows.   We slice the five-dimensional geometry onto each brane, and show that the induced exterior on our brane has the usual Schwarzschild/Vaidya monopole at leading order, with subleading Weyl/Kaluza–Klein corrections in the localized-feeding branch. We construct a perturbative gradient expansion solution satisfying the regular bulk equations and brane junction conditions.  The matter sector satisfies the null energy condition for positive mass growth.  The linear perturbation stability is investigated, and for supermassive seeds with horizon radius far larger than the interbrane scale, the dangerous long-wavelength mode is absent.  The primary observational consequences are overmassive high-redshift black holes in underdeveloped hosts, hidden mass growth relative to the luminous accretion budget, LISA-band heavy-seed mergers, and the absence of primordial fossils required by PBH explanations.
\end{abstract}

\tableofcontents
\newpage

\section{Introduction}

The origin of the first supermassive black holes has become a precision problem rather than a speculative curiosity.  In the standard picture, black holes grow by accreting baryonic gas and by merging with other black holes.  This mechanism certainly operates in galaxies at late times.  The difficulty is the timing.  Observations now point to massive accreting black holes when the Universe was only a few hundred million years old.  Some of these systems appear to contain black holes that are already too massive for their visible hosts, or at least too massive to be comfortably explained by ordinary stellar remnants growing through standard radiatively efficient accretion.  The tension is not merely that a large mass is present; it is that the required baryonic supply, radiative output, feedback history, and host-galaxy assembly must all occur very early and remain mutually consistent.

The usual astrophysical routes are powerful but demanding.  Light stellar-remnant seeds can in principle reach large masses if they form early and accrete nearly continuously, but this requires favorable duty cycles and limited feedback.  Heavy direct-collapse seeds start closer to the required mass scale, but they need special environments: rapid gas inflow, suppression of fragmentation, efficient angular-momentum transport, and often a strong Lyman--Werner background.  These channels remain important and may explain part of the population, but they do not obviously produce every overmassive high-redshift object without some tuning of environment or duty cycle.  Reviews of the assembly of the first massive black holes and the direct-collapse scenario emphasize both the promise and the difficulty of these routes \cite{Volonteri:2010,Inayoshi:2019fun,Smith:2017fgi,Bromm:2002hb,Begelman:2006db,Lodato:2006}.

Primordial black holes provide a conceptually different possibility.  If sufficiently massive primordial black holes were produced in the early Universe, they could act as seeds before any stars formed.  The strength of this explanation is that it bypasses the need for early baryonic collapse.  Its cost is that primordial black holes are not just isolated astrophysical objects; their formation mechanism generally leaves early-Universe fossils.  Depending on the model, these fossils can include enhanced small-scale curvature perturbations, non-Gaussian tails, induced gravitational waves, phase-transition relics, cosmic-string or domain-wall signatures, and CMB spectral distortions.  The mass range relevant for high-redshift supermassive-black-hole seeds is particularly constrained for Gaussian fluctuation scenarios, because the perturbations required to form such objects can dissipate and distort the microwave background.  For this reason, PBHs are a useful comparison class for the present work but not a uniquely compelling explanation without additional evidence \cite{Carr:2020gox,Sasaki:2018dmp,Nakama:2017xvq}.

The observational motivation has sharpened rapidly.  The quasar J0313--1806 at redshift $z=7.6$ already hosts a black hole with mass of order a billion solar masses \cite{Wang:2021gfh}.  JWST and Chandra have identified candidates such as UHZ1 near redshift ten, where the inferred black-hole mass can be comparable to the host stellar mass \cite{Bogdan:2023,Natarajan:2023rxj,Goulding:2023}.  GN-z11, at redshift $z=10.6$, shows spectroscopic evidence for accretion onto a massive black hole \cite{Maiolino:2023thp}.  Compact red JWST sources, often called little red dots, have further complicated the picture.  Some may be dense stellar systems, some may be obscured accreting black holes, and some may be composites.  Their interpretation remains under active debate, but the very existence of plausible early compact AGN candidates makes it timely to explore nonstandard seed channels \cite{Rusakov:2025}.

This paper develops one such channel.  We consider a brane-world setup in which our visible universe is a brane embedded in a five-dimensional bulk, and a second donor brane contains a hidden compact object or dense donor sector.  Gravity propagates through the bulk, while ordinary matter on the two branes need not share Standard Model gauge interactions.  If a sufficiently compact donor-side object forms a common five-dimensional horizon, then our brane may intersect the same horizon.  An observer on our brane sees a black hole.  Matter falling into the donor-side portion of the horizon increases the common horizon mass and therefore increases the black-hole mass inferred on our brane.  The donor matter does not emerge into our exterior universe.  It enters the common interior.  The mechanism therefore transports gravitational mass, not visible matter.

This distinction is the central idea.  A common horizon is not a portal through which gas flows out into our galaxy.  It is a higher-dimensional black object whose induced exterior on our brane is Schwarzschild-like, or Vaidya-like when it grows.  If gas, stars, broad-line clouds, or photons on our brane orbit this object, they infer an ordinary black-hole mass.  They cannot tell from the exterior whether part of the mass originally entered through another brane.  In this sense the model provides a hidden mass-growth channel rather than a hidden matter-injection channel.  That feature is essential for cosmology, because it avoids dumping a large homogeneous radiation or baryon component into our plasma.

The gravitational setting is well motivated by previous work, though the high-redshift seed application appears to be new.  Randall--Sundrum models show how a brane embedded in a warped anti-de Sitter bulk can recover four-dimensional gravity at long distances \cite{Randall:1999ee,Randall:1999vf}.  The effective Einstein equations on a brane include a projected Weyl term that encodes nonlocal bulk gravity \cite{Shiromizu:1999wj,Maartens:2010ar}.  Brane-world black holes, black strings, black cigars, and tidal-charge metrics have been studied for decades \cite{Chamblin:1999by,Dadhich:2000am,Kanti:2004nr}.  Separately, black funnels in anti-de Sitter gravity provide a language for connected horizons and horizon-mediated transport \cite{Hubeny:2009ru,Santos:2012he,Fischetti:2012ps,Emparan:2013xia}.  Our proposal takes the connected-horizon idea and applies it to early black-hole growth: the donor brane supplies energy to a common horizon, and our brane sees the resulting gravitational mass as a black-hole seed.

We adopt the simplest geometry that can be analyzed exactly,  which is a warped Schwarzschild black string. The spatial cross-section of the horizon is a connected tube stretching between the two branes.  When the metric is sliced at the location of our brane, the induced exterior is the usual four-dimensional black-hole form with a mass parameter determined by the common horizon.  When donor-side matter falls in, the mass parameter becomes time dependent.  The five-dimensional Einstein equations then require a positive null stress tensor, exactly as in ordinary Vaidya accretion, and the null energy condition is satisfied for positive mass growth. 

The resulting phenomenology is distinctive but not a single-object smoking gun.  A black-string seed would look like an ordinary black hole in its local exterior gravitational field.  The evidence would come from population-level and multi-messenger consistency: black holes that are too massive for their hosts, hidden growth relative to the luminous accretion budget, LISA-band mergers of heavy seeds, compatibility with pulsar-timing-array descendants, and the absence of the primordial fossils expected from PBH explanations.  The model is therefore falsifiable in several ways.  It fails if the required high-redshift masses disappear with better data, if ordinary accretion and obscuration explain the demographics, if LISA sees no heavy seeds where the electromagnetic population requires them, or if a PBH formation fossil is detected at the relevant mass scale.

The paper is organized as follows.  Section \ref{sec:model} defines the two-brane common-horizon geometry, includes a color illustration of the physical picture, and derives the induced black-hole mass and potential seen on our brane.  Section \ref{sec:einstein} solves the five-dimensional Einstein equations for the static and accreting common-horizon branch, including the brane stress tensors, bulk cosmological constant, and accreting null matter.  Section \ref{sec:obs} develops the main observational signatures across high-redshift astrophysics, cosmology, and gravitational waves.  Section \ref{sec:disc} summarizes the viable parameter logic, the open calculations, and the clearest falsifiers.

\section{A common horizon that transports gravitational mass}\label{sec:model}

We consider two codimension-one branes, denoted $A$ and $B$, in a five-dimensional bulk.  The brane $A$ is our brane, with coordinates $x^\mu=(t,r,\theta,\phi)$ and position $y=y_A$.  The brane $B$ is a hidden donor brane at $y=y_B$.  Matter fields on each brane may be localized to their own brane, while gravity propagates in the five-dimensional bulk.  To avoid an overlap in notation, throughout the rest of the paper the labels $A$ and $B$ refer only to the two branes.  The warp exponent is denoted by $W(y)$, with $W_i\equiv W(y_i)$, and bulk tensor indices will be written as $M,N=0,1,2,3,4$.

The key object is a common event horizon whose spatial cross-section connects the two branes.  In the simplest analytic branch the common horizon is the horizon of a warped black string.  A fully localized black string can be viewed as a deformation of this branch in which the horizon radius varies  across the bulk, but it already captures the essential point: the two brane black holes are not two separate objects connected by a matter pipe.  They are two slices of one higher-dimensional trapped region.

\begin{figure}[t]
\centering
\includegraphics[width=0.96\textwidth]{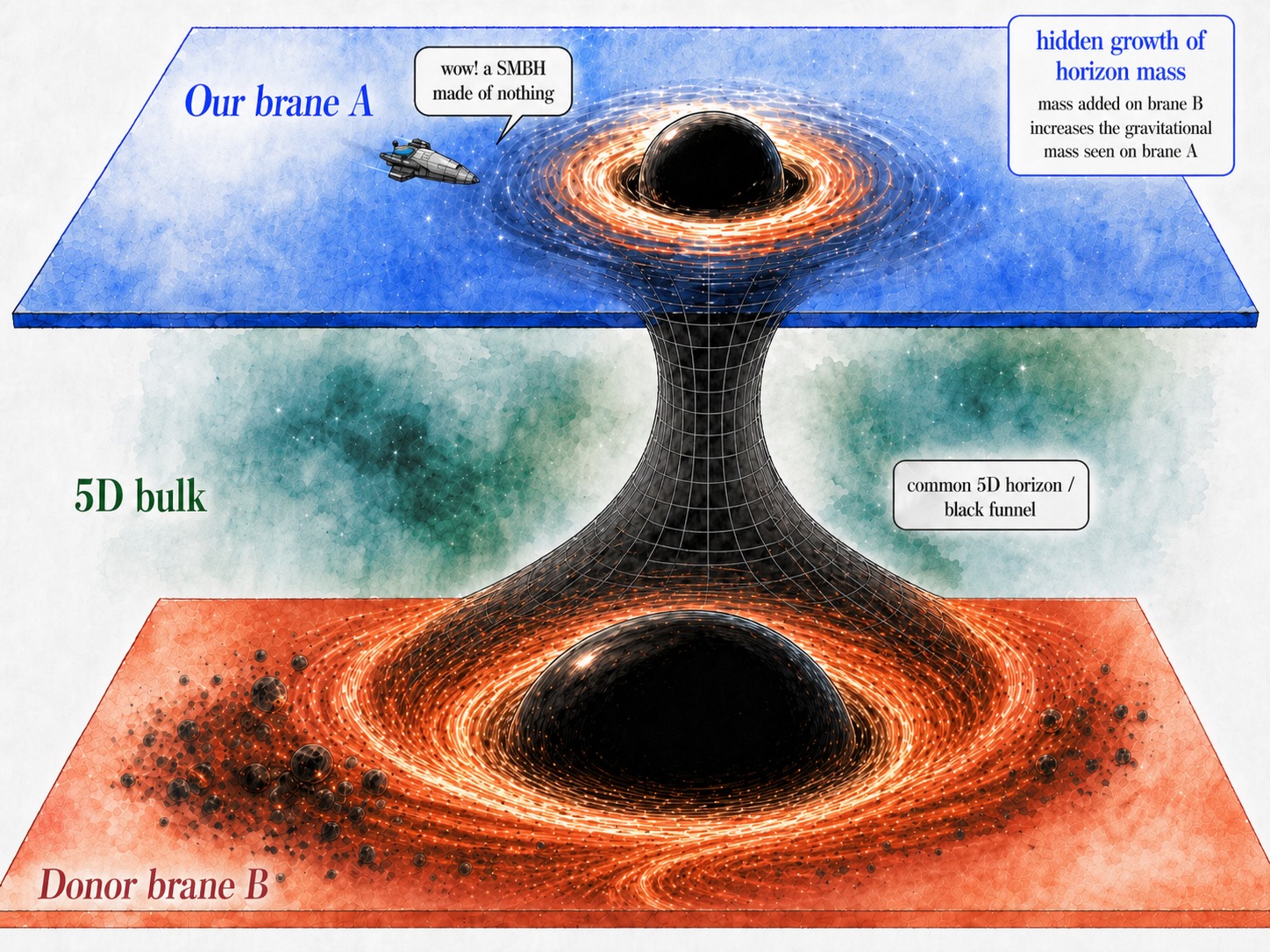}
\vspace{0.5ex}
\parbox{0.94\textwidth}{\small\textbf{Figure 1.} Color-media illustration of the common-horizon model.  The upper blue sheet is our brane $A$, the lower red sheet is a hidden donor brane $B$, and the central dark structure is one connected five-dimensional horizon.  The donor-side black hole and accreting matter increase the same horizon mass that appears on brane $A$ as a visible SMBH seed.  The mechanism therefore transports gravitational mass through the common horizon, not visible matter into the exterior of our brane.}
\end{figure}

The local metric we have adopted is
\begin{equation}
  \dd s_5^2=e^{-2W(y)}\left[-f(r)c^2\dd t^2+\frac{\dd r^2}{f(r)}+r^2\dd\Omega_2^2\right]+\dd y^2,
  \qquad f(r)=1-\frac{2m}{r}.
  \label{eq:staticmetric}
\end{equation}
Here $m$ is a length parameter, $\ell$  denotes the curvature scale, and $W(y)=|y|/\ell$.  In our current work, we do not acquire the large hierarchy arising from the warp factor between 2 branes. 

Let's Introduce the physical time and radius coordinates for observers living on the brane at fixed position $y_i$,
\begin{equation}
  T_i=e^{-W_i}t,
  \qquad R_i=e^{-W_i}r .
\end{equation}
Then the induced metric becomes
\begin{equation}
  \dd s_i^2=-\left(1-\frac{2m e^{-W_i}}{R_i}\right)c^2\dd T_i^2
  +\left(1-\frac{2m e^{-W_i}}{R_i}\right)^{-1}\dd R_i^2+R_i^2\dd\Omega_2^2 .
  \label{eq:slicedmetric}
\end{equation}
Thus each brane observer sees a Schwarzschild exterior with horizon radius
\begin{equation}
  R_{h,i}=2m e^{-W_i}.
\end{equation}
Defining the gravitational mass inferred on brane $i$ by $R_{h,i}\equiv \frac{2G_iM_i}{c^2}$,
where $G_i$ is the Newton constant measured on that brane, gives
\begin{equation}
  M_i=\frac{c^2m e^{-W_i}}{G_i}.
  \label{eq:massslice}
\end{equation}
The physical meaning is direct: the exterior mass inferred by stars, gas, broad-line clouds, lensing, or Eddington arguments on the brane $A$ is the common-horizon mass parameter sliced through brane $A$.  It does not reveal whether the energy originally entered through the brane $A$, the brane $B$, or the bulk.

The horizon is the hypersurface  $r=2m$.
The normal to $r-2m=0$ has norm $g^{rr}=e^{2W}f(r)$, so it is null at $r=2m$.  A spatial cross-section of the horizon has metric
\begin{equation}
  \dd s_H^2=e^{-2W(y)}(2m)^2\dd\Omega_2^2+\dd y^2.
  \label{eq:horizonmetric}
\end{equation}
On the interval $y_A\le y\le y_B$ this has topology $S^2\times I$.   The proper areal radius of that same horizon can differ from slice to slice, $R_h(y)=2m e^{-W(y)}$, so the donor-side horizon can be geometrically larger or smaller than the horizon inferred by observers on our brane. From the viewpoint of an exterior observer on brane $A$, a high-redshift SMBH seed produced by this channel, despite its enormous mass, there is probably literally ``nothing'' behind the horizon on our brane.   Nothing made of our-brane baryons had to collapse in the visible host, no donor atoms fly out of the bulk, and behind the $A$-brane horizon there need not be a conventional four-dimensional interior filled with our-sector matter.  What exists is a five-dimensional trapped region whose gravitational charge intersects our brane.  Astronomers on brane $A$ can weigh the object, watch gas orbit it, and measure its lensing field; they cannot open the horizon and inspect a ledger of where the mass entered.  In this sense the mass is real, while the local visible construction history can be absent.

Assuming that we can neglect the mass accretion effect on brane A, as well as the possible energy loss due to the gravitational waves radiation, matter falling through the donor-side portion of the common horizon increases $m$ in the eq. (\ref{eq:staticmetric}), and the A-brane mass growth law is 
\begin{equation}
\frac{\dd M_A}{\dd T_A}=e^{-2(W_B-W_A)}\frac{\dd M_B}{\dd T_B},
  \label{eq:growthlaw}
\end{equation}
where we have used the relation between the effective Newtonian constants on two branes to get this relative growth law, namely $\frac{G_B}{G_A}=e^{-2(W_B-W_A)}$. This is why the channel is hidden.  Donor particles do not cross into the exterior of brane $A$; they enter the common interior and change the mass parameter that controls the $A$-brane exterior geometry.  The effect is therefore closer to horizon bookkeeping than to transport of ordinary matter through a wormhole-like gate. The black hole seen by an exterior observer on brane A becomes overmassive relative to the stellar mass and luminous accretion history of its $A$-brane host.  This is the core mechanism.  It explains how a genuine gravitational mass can appear early without requiring the visible baryonic environment to have built it.  The host can therefore look young, gas-poor, or underassembled while the central gravitational mass already looks mature.

Our model naturally explains not only the excess mass itself, but also its emergence at sufficiently early times in the evolution of the universe. In general, we do not expect the two branes to share the same cosmic history. The donor brane may possess its own localized stress tensor, reheating temperature, particle content, cooling dynamics, perturbation spectrum, compact-object abundance, and merger history. In this generic asymmetric scenario, the donor sector can form—or more efficiently sustain—compact horizons earlier than the visible sector.

The weak-field potential inferred by an $A$-brane observer follows immediately from Eq.~\eqref{eq:slicedmetric}.  At long distance regime $R_A\gg R_{h,A}$,

\begin{equation}
  \Phi_A(R_A,T_A)\simeq -\frac{G_AM_A(T_A)}{R_A}.
  \label{eq:potentialA}
\end{equation}
The leading exterior potential is therefore indistinguishable from that of an ordinary black hole of mass $M_A$.  Subleading Kaluza--Klein, Weyl/tidal-charge, and multipole corrections may appear in a localized nonuniform black string, but the monopole mass is fixed by the common horizon.  This is both a strength and a limitation of the model: a single well-resolved orbit near the exterior Schwarzschild region would usually measure the mass, not the provenance of that mass.

This distinction reconciles two different gravitational regimes.  If matter on brane $B$ remains outside any common horizon, brane $A$ feels it as ``shadow matter'' through the bulk Green's function, with a schematic potential
\begin{equation}
  \Phi_A^{\rm shadow}(R)=-\frac{G_{AB}^{\rm cross}M_B}{R}+\hbox{KK/radion/Weyl corrections},
\end{equation}
where $G_{AB}^{\rm cross}$ is the cross-brane gravitational coupling.  Garriga and Tanaka showed in a two-brane RS setup that matter on the other brane gravitates on our brane, with model-dependent lensing differences \cite{Garriga:1999yh}.  Once that mass is inside a common horizon, however, the correct exterior description on brane $A$ is Eq.~\eqref{eq:potentialA}.  The mass has become common-horizon mass rather than external shadow matter.  The difference is analogous to the difference between orbiting a distant dark companion and orbiting a black hole whose horizon has already swallowed that companion: the exterior monopole may be the same, but the causal bookkeeping is not.

\section{5D Einstein equations and null dust accretion solution}\label{sec:einstein}
In this section, we first explicitly demonstrate that the static metric of the common-horizon sector, given by eq. (\ref{eq:staticmetric}), satisfies the five-dimensional Einstein equations. We then extend this metric to a time-dependent feeding solution, and finally analyze the stability of these solutions.

\subsection{Static metric, stabilization, and effective Newton constants}
\label{sec:static-metric-stabilized}

The bulk coordinates are denoted by $X^M$, with capital Latin indices
$M,N=0,1,2,3,4$.  The two branes are denoted by $A$ and $B$.
Each brane is a codimension-one hypersurface $\Sigma_i$, with
$i=A,B$.  Brane-intrinsic coordinates are denoted by $x^\mu$, with
Greek indices $\mu,\nu=0,1,2,3$.  The five-dimensional bulk manifold is
denoted by $\mathcal M$, the five-dimensional metric by $g_{MN}$, and
the metric induced on brane $i$ by $h^{(i)}_{\mu\nu}$.  We use
$\kappa_5^2=8\pi G_5$, where $G_5$ is the five-dimensional Newton
constant.

The action, including the Gibbons--Hawking--York boundary terms needed
for a well-defined variational principle, is
\begin{equation}
\begin{split}
  S={}&
  \frac{1}{2\kappa_5^2}
  \int_{\mathcal M}\dd^5X\sqrt{-g}\,(R-2\Lambda_5)
  +\frac{1}{\kappa_5^2}
  \sum_{i=A,B}\int_{\Sigma_i}\dd^4x\sqrt{-h_i}\,K_i
  \\
  &-\sum_{i=A,B}\lambda_i\int_{\Sigma_i}\dd^4x\sqrt{-h_i}
  +S_{\rm GW}+S_A+S_B .
\end{split}
  \label{eq:fullaction}
\end{equation}
Here $R$ is the five-dimensional Ricci scalar, $\Lambda_5$ is the bulk
cosmological constant, $\lambda_i$ is the tension of brane $i$, and
$K_i$ is the trace of the extrinsic curvature
$K^{(i)}_{\mu\nu}$ of $\Sigma_i$.  The sign of each boundary term is
fixed by the corresponding outward-normal convention.  The actions
$S_A$ and $S_B$ contain non-gravitational matter localized on branes
$A$ and $B$, respectively.  The symbol $S_{\rm GW}$ denotes the
Goldberger--Wise stabilization sector, whose dynamics is briefly summarized in the appendix section \ref{GW} . 

Varying Eq.~\eqref{eq:fullaction} gives the distributional field
equation
\begin{equation}
  G_{MN}+\Lambda_5g_{MN}
  =
  \kappa_5^2T^{\rm GW}_{MN}
  +
  \kappa_5^2
  \sum_{i=A,B}
  \left[
    -\lambda_i h^{(i)}_{MN}
    +\tau^{(i)}_{MN}
  \right]\delta_i .
  \label{eq:fullEinstein}
\end{equation}
Here $G_{MN}$ is the five-dimensional Einstein tensor,
$T^{\rm GW}_{MN}$ is the bulk stress tensor of the Goldberger--Wise
scalar, $\tau^{(i)}_{MN}$ is the non-tension stress tensor localized on
brane $i$, and $\delta_i$ is the properly normalized delta function
with support on $\Sigma_i$.  In the small-backreaction approximation,
$T^{\rm GW}_{MN}$ fixes the modulus but does not significantly deform
the exterior black-string geometry.

For the RS-tuned exterior background, take
\begin{equation}
  \Lambda_5=-\frac{6}{\ell^2},
  \qquad
  W(y)=\frac{y}{\ell},
  \qquad
  0\le y\le d .
  \label{eq:RSwarp}
\end{equation}
We place brane $A$ at $y_A=0$ and brane $B$ at $y_B=d$.  The tuned
brane tensions are
\begin{equation}
  \lambda_A=+\frac{6}{\kappa_5^2\ell},
  \qquad
  \lambda_B=-\frac{6}{\kappa_5^2\ell}.
  \label{eq:RStensionsAB}
\end{equation}
Thus brane $B$ is the negative-tension brane in this benchmark.

Away from brane delta functions, the metric has warped-product form
\begin{equation}
  \dd s_5^2
  =
  e^{-2W(y)}
  \hat g_{\mu\nu}(x)\dd x^\mu\dd x^\nu
  +
  \dd y^2 .
  \label{eq:warpedproductmetric}
\end{equation}
A hat denotes a quantity built from the four-dimensional metric
$\hat g_{\mu\nu}$.  For this metric,
\begin{align}
  R_{\mu\nu}^{(5)}
  &=
  \hat R_{\mu\nu}^{(4)}
  +
  \left(W''-4W'^2\right)g_{\mu\nu},
  \label{eq:ricci1}
  \\
  R_{yy}^{(5)}
  &=
  4W''-4W'^2,
  \label{eq:ricci2}
  \\
  R_{\mu y}^{(5)}
  &=
  0,
  \label{eq:ricci3}
  \\
  R^{(5)}
  &=
  e^{2W}\hat R^{(4)}
  +
  8W''-20W'^2 .
  \label{eq:ricci4}
\end{align}
Here $\hat R_{\mu\nu}^{(4)}$ and $\hat R^{(4)}$ are the Ricci tensor
and Ricci scalar of $\hat g_{\mu\nu}$.  If $\hat g_{\mu\nu}$ is the
Schwarzschild metric appearing in Eq.~\eqref{eq:staticmetric}, then
\begin{equation}
  \hat R_{\mu\nu}^{(4)}=0,
  \qquad
  \hat R^{(4)}=0 .
\end{equation}
In the open bulk, $W'=1/\ell$ and $W''=0$, so
\begin{equation}
  R_{MN}^{(5)}
  =
  -\frac{4}{\ell^2}g_{MN},
  \qquad
  R^{(5)}
  =
  -\frac{20}{\ell^2}.
\end{equation}
Therefore
\begin{equation}
  G_{MN}^{(5)}
  =
  \frac{6}{\ell^2}g_{MN},
  \qquad
  G_{MN}^{(5)}+\Lambda_5g_{MN}=0 .
\end{equation}
Thus the static warped Schwarzschild metric in Eq.~\eqref{eq:staticmetric}
solves the regular five-dimensional Einstein equations away from the
branes, up to the intentionally neglected small Goldberger--Wise
backreaction.  The brane delta functions are supplied by the jumps of
the extrinsic curvature and are matched by Eq.~\eqref{eq:RStensionsAB},
with small retunings if the stabilization sector has non-negligible
vacuum energy.

The effective four-dimensional Newton constants follow from the
zero-mode graviton kinetic term.  We define the reduced Planck mass on
brane $i$ by
\begin{equation}
  M_{i,\rm Pl}^2\equiv\frac{1}{8\pi G_i},
\end{equation}
where $G_i$ is the Newton constant measured by observers using the
physical metric on brane $i$.  In the physical frame of brane $i$,
\begin{equation}
  M_{i,\rm Pl}^2
  =
  \frac{e^{2W_i}}{\kappa_5^2}
  \int_{\rm orb}\dd y\,e^{-2W(y)},
  \qquad
  W_i\equiv W(y_i).
  \label{eq:Planckgeneral}
\end{equation}
The integral is over the orbifold covering space; equivalently one may
use an interval convention with the corresponding orbifold-normalized
$\kappa_5^2$.

For $W_A=0$ and $W_B=d/\ell$, Eq.~\eqref{eq:Planckgeneral} gives
\begin{equation}
  M_{A,\rm Pl}^2
  =
  \frac{\ell}{\kappa_5^2}
  \left(1-e^{-2d/\ell}\right),
  \qquad
  8\pi G_A
  =
  \frac{\kappa_5^2}
  {\ell\left(1-e^{-2d/\ell}\right)} ,
  \label{eq:GAcompact}
\end{equation}
and
\begin{equation}
  M_{B,\rm Pl}^2
  =
  \frac{\ell}{\kappa_5^2}
  \left(e^{2d/\ell}-1\right),
  \qquad
  8\pi G_B
  =
  \frac{\kappa_5^2}
  {\ell\left(e^{2d/\ell}-1\right)} .
  \label{eq:GBcompact}
\end{equation}
Therefore
\begin{equation}
  G_A>0,
  \qquad
  G_B>0,
  \label{eq:positiveGbothbranes}
\end{equation}
even though $\lambda_B<0$.  The sign of the local brane tension is not
the sign of the measured zero-mode Newton constant in the compact
stabilized theory.  Without radion stabilization, the long-distance
theory is generally scalar-tensor rather than pure Einstein gravity;
the assumption \eqref{eq:radionhealthy} is what allows the tensor
Newton constant above to be used as the leading coupling. The ratio of the two Newton constants is
\begin{equation}
  \frac{G_B}{G_A}
  =
  e^{-2d/\ell}
  =
  e^{-2(W_B-W_A)} .
  \label{eq:GratioAB}
\end{equation}
This is the relation used in the common-horizon mass bookkeeping eq. (\ref{eq:growthlaw}).

In passing, we 
shall discuss stability in the regime relevant for SMBH seeds.  
For a compact interval of proper size $d$ between the branes, the allowed fifth-dimensional wave numbers are schematically
\begin{equation}
  k_n\simeq \frac{n\pi}{d},\qquad n=1,2,\ldots,
\end{equation}
up to warp and boundary-condition corrections.  The Gregory--Laflamme instability of a five-dimensional Schwarzschild black string exists only for wavelengths longer than a critical value, usually written
\begin{equation}
  kR_h < k_{\rm GL}R_h\sim\mathcal{O}(1),
\end{equation}
where $R_h$ is the relevant proper horizon radius and $k_{\rm GL}$ is the critical GL wave number \cite{Gregory:1993vy}.  In a warped geometry the proper horizon radius varies across the interval,
\begin{equation}
  R_h(y)=2m e^{-W(y)},\qquad R_{h,\min}=\min_{y\in[y_A,y_B]} R_h(y).
\end{equation}
A compact extra dimension removes the dangerous mode if the lowest nonzero mode satisfies
\begin{equation}
  \frac{\pi}{d} > \frac{\mathcal{O}(1)}{R_{h,\min}}.
  \label{eq:GLstable}
\end{equation}
For an SMBH seed with $M_A\simeq10^6\Msun$, the horizon radius seen on our brane is $R_{h,A}=2G_AM_A/c^2\sim \mathcal{O}(10^9)\,\mathrm{m}$.  If the warp variation across the interval is not exponentially large, $R_{h,\min}$ is of the same order, while a microscopic interbrane scale may be $d\lesssim10^{-4}\,\mathrm{m}$.  Then $\pi/d$ exceeds $1/R_{h,\min}$ by more than ten orders of magnitude.  Physically, the horizon is extremely fat compared with the extra dimension, so the long-wavelength necking mode that destabilizes an infinite string is not available. 
It is worth mentioning that if a model uses a very large warp hierarchy, instability issues may arise.  Thus our solution is perturbatively stable in the sector that can be treated analytically: no available GL mode in the supermassive compact-interval regime, and a heavy radion suppressing brane-separation instabilities.  A fully localized black string still deserves numerical study, but the controlled limit is internally consistent and gives a well-defined target for such simulations.

\subsection{Localized $B$-brane null dust accretion}
\label{sec:Bbrane-null-accretion-v3}

We now model donor-side feeding by a null stress tensor localized on brane B. The construction presented below provides a controlled long-wavelength solution of the five-dimensional Einstein equations that simultaneously satisfies the bulk field equations and the junction conditions on both branes at first order in the gradient expansion. It should be regarded as a perturbative description of a weakly nonuniform, accreting common horizon rather than an exact closed-form nonlinear black-string solution. Nevertheless, within its regime of validity, it captures the leading gravitational response of the common horizon to donor-brane accretion and provides a self-consistent framework for studying the induced mass growth observed on brane A.

Let $Y$ be a Gaussian-normal coordinate transverse to brane $B$, and
define the $Z_2$-even proper distance
\begin{equation}
Y\equiv y-y_B,\qquad  z\equiv |Y|.
  \label{eq:Bnullv3-zdef}
\end{equation}
The absolute value is essential.  The metric is continuous across $Y=0$, while its first normal derivative jumps; that jump is precisely what produces the distributional brane stress tensor.  On a one-sided physical interval, $z$ is simply the proper distance from brane $B$, with the same condition imposed as an Israel boundary condition. On the one-sided compact interval, $0\le z\le d$, brane $B$ sits at
$z=0$ and the second brane sits at $z=d$.  We keep the donor tension
sign explicit:
\begin{equation}
  \epsilon_B
  \equiv
  \frac{\kappa_5^2\lambda_B\ell}{6}
  =
  \pm1,
  \qquad
  a_B(z)=e^{-\epsilon_B z/\ell}.
  \label{eq:Bnullv3-warp}
\end{equation}

Let $x^a=(V,R,\theta,\phi)$ be coordinates intrinsic to brane $B$,
with lower Latin indices $a,b=0,1,2,3$ used for brane-$B$ tensors in
this subsection.  The coordinate $V$ is an ingoing advanced coordinate
with dimensions of length. The coordinate $R$ is the physical areal radius on
brane $B$.  The induced donor-brane metric which describing null dust accretion on the brane B is
\begin{equation}
  \dd s_B^2
  =
  h^{(B)}_{ab}\dd x^a\dd x^b
  =
  -F_B(V,R)\dd V^2
  +2\dd V\dd R
  +R^2\dd\Omega_2^2,
  \qquad
  F_B(V,R)=1-\frac{2m_B(V)}{R}.
  \label{eq:Bnullv3-induced}
\end{equation}
Here $\dd\Omega_2^2$ is the unit two-sphere metric and
\begin{equation}
  m_B(V)=\frac{G_BM_B(V)}{c^2}
\end{equation}
is the length-valued Vaidya mass parameter inferred by observers on
brane $B$.  The Ricci tensor reads
\begin{equation}
  R^{(B)}_{ab}
  =
  \frac{2\dot m_B}{R^2}k_ak_b,
  \qquad
  k_a=-\partial_aV,
  \qquad
  h_{(B)}^{ab}k_ak_b=0 .
  \label{eq:Bnullv3-kdef}
\end{equation}
The one-form $k_a$ is the ingoing null direction of the Vaidya flow, and $  \dot m_B\equiv\frac{\dd m_B}{\dd V}.$ The energy momentum tensor of the null dust on the brane B is 
\begin{eqnarray}
S_{ab}^{\text{null dust}}=\psi_Bk_ak_b,\qquad
  \psi_B(V,R)
  =
  \frac{\dot m_B(V)}{4\pi G_B R^2}.
\end{eqnarray}

We now extend the brane metric eq. (\ref{eq:Bnullv3-induced}) to the bulk metric in a perturbative manner.  Let $L_4$ denote the
four-dimensional curvature scale of the induced Vaidya geometry, and
assume
\begin{equation}
  \frac{\ell}{L_4}\ll1 .
\end{equation}
This is a very good approximation in our current work, because typically the $\ell$ is experimentally constrained to be at most at micrometer level, while $L_4$ is typically around million or even billion kilometers for SMBHs. The metric on a constant-$Y$ slice is denoted by $q_{ab}(z,x)$.  At
first gradient order, the only trace-free curvature tensor available
on the donor brane is $R^{(B)}_{ab}$, so write
\begin{equation}
  q_{ab}(z,x)
  =
  a_B^2(z)h^{(B)}_{ab}(x)
  +
  f(z)R^{(B)}_{ab}(x)
  +
  \mathcal O\!\left(\frac{\ell^4}{L_4^4}\right).
  \label{eq:Bnullv3-tensoransatz}
\end{equation}
The function $f(z)$ has dimensions of length squared.  Since
$R^{(B)}_{ab}=O(L_4^{-2})$, the correction $fR^{(B)}_{ab}$ is
dimensionless and perturbative when $|f|/L_4^2\ll1$. A direct Gauss-normal calculation gives, away from the branes,
\begin{equation}
  R^{(5)}_{ab}
  +
  \frac{4}{\ell^2}q_{ab}
  =
  \left[
    1-\frac12 f''
    +\frac{2f}{\ell^2}
  \right]R^{(B)}_{ab}
  +
  \mathcal O\!\left(\frac{\ell^2}{L_4^4}\right).
  \label{eq:Bnullv3-bulkodepre}
\end{equation}
Here the prime denotes differentiation with respect to $z$.  Therefore
the regular bulk Einstein equations are solved through all
$O(L_4^{-2})$ terms if and only if
\begin{equation}
  f''-\frac{4}{\ell^2}f=2 .
  \label{eq:Bnullv3-fode}
\end{equation}

The induced metric on brane $B$ must be exactly
Eq.~\eqref{eq:Bnullv3-induced}.  Since $a_B(0)=1$, this imposes
\begin{equation}
  f(0)=0 .
  \label{eq:Bnullv3-f0}
\end{equation}
The null part of the brane-$B$ Israel condition gives
\begin{equation}
  \frac12f'(0)R^{(B)}_{ab}
  =
  -\frac{\kappa_5^2}{2}\psi_Bk_ak_b .
\end{equation}
One obtains
\begin{equation}
  f'(0)
  =
  -\frac{\kappa_5^2}{8\pi G_B}
  \equiv
  -\beta_B\ell,
  \qquad
  \beta_B\equiv
  \frac{\kappa_5^2}{8\pi G_B\ell}.
  \label{eq:Bnullv3-betadef}
\end{equation}
The dimensionless parameter $\beta_B$ encodes the effective Newton
coupling on the donor brane. Solving Eq.~\eqref{eq:Bnullv3-fode} with
Eqs.~\eqref{eq:Bnullv3-f0} and \eqref{eq:Bnullv3-betadef} gives
\begin{equation}
  f(z)
  =
  -\frac{\ell^2}{2}
  +
  \frac{\ell^2}{4}(1-\beta_B)e^{2z/\ell}
  +
  \frac{\ell^2}{4}(1+\beta_B)e^{-2z/\ell}.
  \label{eq:Bnullv3-fsolution-general}
\end{equation}
The homogeneous terms in Eq.~\eqref{eq:Bnullv3-fsolution-general} are
fixed by the second brane.  Let the second brane sit at $z=d$ and carry
only its tuned tension, $  \lambda_A=-\lambda_B $. 
With inward normal $-\partial_z$ at the second boundary, the pure-tension
Israel condition at $z=d$ gives the Robin condition
\begin{equation}
  f'(d)+\frac{2\epsilon_B}{\ell}f(d)=0 .
  \label{eq:Bnullv3-Arobin}
\end{equation}
Substituting Eq.~\eqref{eq:Bnullv3-fsolution-general} into
Eq.~\eqref{eq:Bnullv3-Arobin} gives
\begin{equation}
  \beta_B
  =
  \epsilon_B\left(1-e^{-2\epsilon_B d/\ell}\right).
  \label{eq:Bnullv3-betacompact}
\end{equation}
Combining this with Eq.~\eqref{eq:Bnullv3-betadef} reproduces the
compact two-brane Newton constant,
\begin{equation}
  8\pi G_B
  =
  \frac{\kappa_5^2}
  {\ell\,\epsilon_B\left(1-e^{-2\epsilon_B d/\ell}\right)},
  \label{eq:Bnullv3-GBcompact}
\end{equation}
which is always positive regardless of the brane tension. Thus positive accretion on a negative-tension brane is obtained only
after the compact second boundary condition is imposed.

Given the eq. (\ref{eq:Bnullv3-tensoransatz}), now we can write down the 5D metric, 
\begin{equation}
  \dd s_5^2
  =
  \dd Y^2
  -
  \left[
    a_B^2F_B
    -
    \frac{2\dot m_B}{R^2}f(z)
  \right]\dd V^2
  +2a_B^2\dd V\dd R
  +a_B^2R^2\dd\Omega_2^2
  +
  \mathcal O\!\left(\frac{\ell^4}{L_4^4}\right).
  \label{eq:Bnullv3-explicitmetric}
\end{equation}
At $z=0$, $a_B=1$ and $f(0)=0$, so the induced metric is exactly the
donor Vaidya metric eq. (\ref{eq:Bnullv3-induced}). The localization of the source on brane $B$ follows from the one-sided
normal derivative, and it can be checked explicitly. The 
singular Einstein tensor reads
\begin{equation}
  G_{MN}+\Lambda_5g_{MN}=\kappa_5^2T_{MN},
\end{equation}
where the localized surface stress tensor on brane $B$ is
\begin{equation}
  T^{(B)}_{MN}
  =
  S^{(B)}_{MN}\delta(Y),
  \qquad
  S^{(B)}_{ab}
  =
  -\lambda_Bh^{(B)}_{ab}
  +
  \psi_Bk_ak_b,
  \qquad
  S^{(B)}_{aY}=S^{(B)}_{YY}=0 .
  \label{eq:Bnullv3-source}
\end{equation}
For any five-dimensional null vector $n^M$, with $\dot{m}_B\ge0$ and thus  $\psi_B\ge0$, we have 
\begin{equation}
  \psi_B k_Mk_N n^M n^N
  =
  \psi_B(k\cdot n)^2
  \ge0.
\end{equation}
Thus the accreting null dust component satisfies the
null energy condition.  If $\lambda_B<0$, the negative brane tension is
a separate background ingredient; the feeding matter itself still has
positive null energy.

  The correction in Eq.~\eqref{eq:Bnullv3-explicitmetric}
is first order in the four-dimensional curvature expansion because
$f=O(\ell^2)$ and $\psi_B\sim\dot m_B/(G_BR^2)$. The regular bulk residual is at the order $  \mathcal O\!\left(\frac{\ell^2}{L_4^4}\right).$
Since the bulk equations contain two derivatives in the transverse direction, $\partial_z^2\sim \ell^{-2}$,  the omitted \(O(\ell^4/L_4^4)\) in the metric eq.(\ref{eq:Bnullv3-explicitmetric}) generate contributions to the Einstein tensor at the order of \(O(\ell^2/L_4^4)\), which are required to cancel the residual regular bulk terms.  Consequently, the metric displayed above should be interpreted as the leading long-wavelength approximation to a localized-accretion bulk solution.

Now let's investigate the growth of the inferred gravitational mass on the brane A. Firstly, 
On brane $B$, the outgoing expansion of
Eq.~\eqref{eq:Bnullv3-induced} vanishes at
\begin{equation}
  R_H^B(V)=2m_B(V).
\end{equation}
Therefore
\begin{equation}
  \frac{\dd R_H^B}{\dd V}
  =
  2\dot m_B
  =
  8\pi G_B\left(R_H^B\right)^2
  \psi_B\!\left(V,R_H^B\right).
  \label{eq:Bnullv3-Bhorizon}
\end{equation}
Since $G_B>0$, positive null energy $\psi_B>0$ is equivalent to
$\dot m_B>0$, and the donor-side horizon section expands, which consequently increases the horizon radius seen by an observer on our brane, because at the leading order we have 
\begin{equation}
  R_H^A(V)
  =
  a_A R_H^B(V)
  +
  \mathcal O\!\left(\frac{\ell^2}{L_4}\right),
  \qquad
  a_A\equiv a_B(d)=e^{-\epsilon_B d/\ell}.
  \label{eq:Bnullv3-Ahorizon}
\end{equation}
The growth of the mass inferred on brane $A$ is therefore 
\begin{equation}
  \frac{\dd M_A}{\dd T_A}
  =
  e^{-2d/\ell}
  \frac{\dd M_B}{\dd T_B}
  +
  \mathcal O\!\left(\frac{\ell^2}{L_4^2}\right),
\end{equation}
where we have adopted the negative-tension donor benchmark, $\epsilon_B=-1$ and
$a_A=e^{d/\ell}$. This is one of main conclusions of this work.   A full nonlinear localized black-string solution would
still require the next-gradient metric, the stabilized radion response
beyond the heavy-radion limit, and the regularity of the connected
horizon across the complete interval, which is beyond the scope of our current work. 

\section{Observational signatures}\label{sec:obs}

The model predicts a black-hole exterior on brane $A$, so the smoking gun is not a visible portal or a stream of matter appearing from nowhere.  The exterior metric is deliberately conservative: at leading order it gives the same monopole force law as an ordinary black hole.  The evidence must therefore come from consistency checks across mass assembly, host-galaxy evolution, and gravitational waves. As this is our first paper on the topic, we do not expect to address all outstanding issues. Instead, in this section, we briefly discuss several potential observational windows through which evidence for, or constraints on, this model may be obtained. We leave more detailed investigations to future work.

One of the most significant observational effects is the overmassive black holes in underdeveloped hosts. The induced potential on brane $A$ is Eq.~\eqref{eq:potentialA}.  Broad-line gas, stars, and photons therefore infer a conventional black-hole mass $M_A$.  If a substantial fraction of the inferred black-hole mass originates from donor-side feeding, the black hole can appear overmassive relative to the visible stellar content of its host. In particular, one expects systems lying significantly above the local $M_{\text{BH}}/M_{\text{host}}$ relation.
This is the most direct astrophysical prediction.  It is also the most immediately testable by JWST, ELT, Roman and future X-ray data.  The caveat is that broad-line virial masses, lensing magnifications, stellar masses, and little-red-dot interpretations are systematics-limited; the model should be tested statistically, not with a single object.  The relevant pattern is an ensemble of central masses that are too large compared with the stellar mass, metallicity, dust content, and star-formation history of their visible hosts.

We also expect a hidden-growth excess.
Let's define the observed black-hole mass-density growth rate $\dot\rho_{\bullet,{\rm obs}}$ and the growth rate supported by visible accretion $\dot\rho_{\bullet,{\rm vis}}$,
A hidden-growth phase can give
\begin{equation}\label{grwothexcess}
\dot\rho_{\bullet,{\rm obs}}\gg\dot\rho_{\bullet,{\rm vis}}.
\end{equation}
Physically, the black hole grows faster than our-brane photons can account for.  This is a cleaner diagnostic than the mass of any individual high-$z$ candidate because it asks whether the integrated mass assembly violates the luminous accretion budget.  Ordinary obscuration, radiative-efficiency evolution, and selection effects must be included before interpreting eq. (\ref{grwothexcess}) as evidence for hidden horizon growth.

Moreover, LISA can be a powerful probe of mergers of heavy seeds.
A binary with source-frame total mass $M$ has observed Schwarzschild-ISCO gravitational-wave frequency
\begin{equation}
  f_{\rm ISCO,obs}\simeq\frac{c^3}{6^{3/2}\pi G_NM(1+z)}
  \simeq4.0\,{\rm mHz}\left(\frac{10^5\Msun}{M}\right)\left(\frac{11}{1+z}\right).
\end{equation}
Thus $10^4$--$10^6\,\Msun$ seed mergers at $z\gtrsim10$ are LISA targets \cite{AmaroSeoane:2017osp}.   A population of high-redshift heavy-seed mergers would strongly favor heavy-seed channels over a purely light stellar-remnant origin.  In the black-string picture, the key extra question is whether the merger rate inferred from gravitational waves matches the electromagnetic count of overmassive high-redshift nuclei without requiring an impossible visible accretion history.

Our model is very much different from the PBH seeds due to the absence of the PBH formation fossils.
A heavy seed does not by itself identify the mechanism.  Inflationary PBHs with $M\sim10^5$--$10^6\,\Msun$ are tied to enhanced curvature perturbations on scales that can produce CMB spectral distortions; Gaussian scenarios in the supermassive range are strongly constrained by $\mu$-distortion limits \cite{Nakama:2017xvq}.  A PIXIE-like mission would sharpen this test \cite{Kogut:2011xw}.  Black-string seeds require no scalar-power spike.  Thus the model is favored only if heavy seeds are observed without the appropriate PBH relics.  Conversely, a clear primordial fossil at the mass scale of the seeds would push the interpretation back toward PBHs or toward a hybrid scenario.

\section{Conclusion and discussion}\label{sec:disc}

We have described a hidden seed mechanism whose central object is  a common five-dimensional horizon which intersects our brane as an ordinary Schwarzschild or Vaidya black hole, while another portion of the same horizon intersects a hidden donor brane.  Donor-side matter that crosses the donor-side horizon increases the common mass parameter.  An exterior observer on brane $A$ therefore infers a growing black hole even when the visible host has not supplied the corresponding baryons.  The mechanism transports gravitational mass through horizon geometry rather than transporting visible matter into the exterior of our brane.

The induced metric on brane $A$ has the standard Schwarzschild monopole, and its Vaidya generalization has the usual positive-energy null flux when the common mass length grows.  Gas clouds, stars, broad-line regions, and photons orbiting far from the horizon simply measure a mass.  They do not measure the brane on which the mass entered the trapped region.  This is the precise sense in which a high-redshift SMBH seed can be ``made of nothing'' from the visible-sector perspective: the object has real gravitational mass, but the local inventory of our-brane material behind its formation can be empty or radically incomplete.

The observational program is therefore not a search for matter appearing from the bulk.  It is a search for an accounting mismatch.  On the electromagnetic side, the model predicts overmassive black holes in underdeveloped hosts and a hidden-growth excess relative to the luminous accretion budget.  On the gravitational-wave side, it predicts heavy-seed mergers in the LISA band if the channel is cosmologically important.   It is most compelling if high-redshift heavy seeds are found without the primordial scalar-power fossils expected from PBH formation at the same mass scale.

The proposal is therefore not a replacement for ordinary astrophysical growth, direct collapse, or PBHs in every object.  It is a sharply defined additional channel: a visible black-hole seed can be the brane slice of a larger higher-dimensional horizon whose feeding history is partly hidden.  The mechanism is unusual, but its exterior limit is calculable, its stability regime is plausible for supermassive seeds, and its observational consequences are concrete.  If early black holes continue to look too massive for their visible histories, the black-string interpretation gives a way to take that mismatch literally: the visible universe may be weighing a horizon whose mass was assembled elsewhere.

\section*{Acknowledgments}
The author thanks G. Domenech, Y. Ma ,  M. Sasaki, T. Tanaka, and G. Zahariade for fruitful discussions.  This work was supported by the grant No. UMO-2021/42/E/ST9/00260 from the National Science Centre, Poland.

\appendix

\section{Goldberger--Wise mechanism}\label{GW}

We take the Goldberger--Wise sector to be a bulk scalar field
$\Phi(X)$ with mass $m_\Phi$ and brane-localized potentials $U_i$:
\begin{equation}
  S_{\rm GW}
  =
  -\frac12
  \int_{\mathcal M}\dd^5X\sqrt{-g}
  \left[
    g^{MN}\partial_M\Phi\partial_N\Phi
    +m_\Phi^2\Phi^2
  \right]
  -\sum_{i=A,B}
  \int_{\Sigma_i}\dd^4x\sqrt{-h_i}\,U_i(\Phi).
  \label{eq:GWaction}
\end{equation}
A simple stiff-potential choice is
\begin{equation}
  U_i(\Phi)
  =
  \frac{\gamma_i}{2}\left(\Phi-v_i\right)^2,
  \qquad
  \gamma_i\ell\gg1 ,
  \label{eq:GWpotentials}
\end{equation}
where $\gamma_i$ are brane-potential stiffness parameters, $v_i$ are
the preferred boundary values of the scalar, and $\ell$ is the AdS
curvature radius.  In the stiff limit,
\begin{equation}
  \Phi(y_A)=v_A,
  \qquad
  \Phi(y_B)=v_B .
  \label{eq:GWboundaryvalues}
\end{equation}
For the background
\begin{equation}
  \dd s_5^2
  =
  \dd y^2+e^{-2y/\ell}\eta_{\mu\nu}\dd x^\mu\dd x^\nu,
  \qquad
  0\le y\le d,
\end{equation}
where $\eta_{\mu\nu}$ is the four-dimensional Minkowski metric and
$d$ is the proper brane separation, the bulk scalar equation is
\begin{equation}
  \Phi''-\frac{4}{\ell}\Phi'-m_\Phi^2\Phi=0.
\end{equation}
A prime denotes differentiation with respect to $y$.  The solution is
\begin{equation}
  \Phi(y)=C_+e^{(2+\nu)y/\ell}+C_-e^{(2-\nu)y/\ell},
  \qquad
  \nu=\sqrt{4+m_\Phi^2\ell^2},
  \label{eq:GWsolution}
\end{equation}
where $C_+$ and $C_-$ are integration constants fixed by the boundary
values.  For $m_\Phi^2\ell^2\ll1$, define
\begin{equation}
  \epsilon_{\rm GW}\equiv \nu-2
  \simeq
  \frac{m_\Phi^2\ell^2}{4}.
\end{equation}
Minimizing the resulting radion potential gives, at leading order in
the small-backreaction limit,
\begin{equation}
  \frac{d}{\ell}
  \simeq
  \frac{1}{\epsilon_{\rm GW}}
  \ln\!\left(\frac{v_A}{v_B}\right),
  \label{eq:GWstabilizedd}
\end{equation}
up to order-one corrections from finite $\gamma_i$ and scalar
backreaction.  The radion is the scalar modulus describing fluctuations
of the inter-brane distance.  We denote its stabilized mass by
$\mu_{\rm rad}$ and assume
\begin{equation}
  \mu_{\rm rad}^2>0,
  \qquad
  \mu_{\rm rad}R_h\gg1,
  \label{eq:radionhealthy}
\end{equation}
where $R_h$ is the characteristic horizon radius of the black object.
Thus the brane separation is fixed on the horizon and accretion
timescales considered below.

\end{document}